\documentclass[conference]{IEEEtran}
\IEEEoverridecommandlockouts

\usepackage{cite}
\usepackage{amsmath,amssymb,amsfonts}
\usepackage{graphicx}
\usepackage{textcomp}
\usepackage{eso-pic}
\usepackage[hidelinks]{hyperref}

\newcommand{\inv}[1]{{#1}^{-1}}

\begin{document}

\AddToShipoutPictureBG*{%
  \AtPageLowerLeft{%
    \put(\LenToUnit{\dimexpr 1in+\hoffset+\oddsidemargin\relax},\LenToUnit{0.4in}){%
      \parbox[b]{\textwidth}{\centering\scriptsize
        \textcopyright~2026 IEEE. Personal use of this material is permitted. Permission from IEEE must be obtained for all other uses, in any current or future media, including reprinting\slash republishing this material for advertising or promotional purposes, creating new collective works, for resale or redistribution to servers or lists, or reuse of any copyrighted component of this work in other works.}}}}

\title{Adaptive Beamforming Using One-Bit Data Samples for Covariance Estimation\\
\thanks{ DISTRIBUTION STATEMENT A. Approved for public release. Distribution is unlimited.
This material is based upon work supported by the Under Secretary of War for Research and Engineering under Air Force Contract No. FA8702-15-D-0001 or FA8702-25-D-B002. Any opinions, findings, conclusions or recommendations expressed in this material are those of the author(s) and do not necessarily reflect the views of the Under Secretary of War for Research and Engineering.
Delivered to the U.S. Government with Unlimited Rights, as defined in DFARS Part 252.227-7013 or 7014 (Feb 2014). Notwithstanding any copyright notice, U.S. Government rights in this work are defined by DFARS 252.227-7013 or DFARS 252.227-7014 as detailed above. Use of this work other than as specifically authorized by the U.S. Government may violate any copyrights that exist in this work.
}
}

\author{
\IEEEauthorblockN{Aaron Kogan, Swaroop Appadwedula}
\IEEEauthorblockA{\textit{Tactical Satellite Communications} \\
\textit{MIT Lincoln Laboratory}\\
Lexington, MA, USA \\
aaron.kogan@ll.mit.edu, swaroop@ll.mit.edu \\
}
}

\maketitle

\begin{abstract}
The use of one-bit quantized complex-valued data samples enables efficient distributed computation of an array data covariance.  This reduced complexity covariance estimate can be used for adaptive beamforming, but requires careful diagonal loading and results in lower performance compared to a full-precision approach.  Two main approaches are presented: a covariance matrix built out of complex one-bit data samples, and a ``hybrid'' cross-covariance between the one-bit data and the full precision data. Signal-to-interference-plus-noise ratio (SINR) estimates for the associated population and finite-sample matrices are derived and compared to the full precision results using tools from linear algebra and statistics. Many strong, well-separated interferers with low overlap with the signal of interest are assumed throughout, none of them dominant.

\end{abstract}

\begin{IEEEkeywords}
Adaptive beamforming, array processing, one-bit quantization, Van Vleck correction
\end{IEEEkeywords}

\section{Introduction}
As cost per antenna element drops, arrays of hundreds of elements become practical, providing many nulling degrees of freedom for adaptive beamforming. However, forming these nulls requires a matrix decomposition, typically QR decomposition of the data matrix, whose cost is linear in the number of samples and quadratic in the number of elements.

One-bit (sign) quantization of complex RF samples has been used in
(a) radio astronomy, where one-bit correlators with the Van Vleck correction \cite{VanVleckMiddleton} are used to compute spatial covariances for direction finding \cite{Benkevitch}, and (b) LMS techniques for adaptive updates where it is called signed regressor \cite{signed_regressor} and sign-sign \cite{sign_sign}. For arrays with computation at each element, an approximate covariance can instead be accumulated from one-bit samples, which are cheap to exchange and multiply, and then factored by Cholesky decomposition, whose cost is independent of the number of samples and below that of QR by more than the ratio of samples to elements.

In this work, we propose two one-bit quantized, data-based covariance estimators and evaluate their signal-to-interference-plus-noise ratio (SINR) performance. For all of the techniques discussed in this paper---unquantized, one-bit, and hybrid---we add diagonal loading in our simulations in order to improve the conditioning of the covariance estimates.  It is known (e.g., \cite{carlson}) that diagonal loading of covariance matrix estimates can improve the robustness of adaptive beamformers and increase associated SINRs, especially with low sample support. While less imperative at moderate sample support for the unquantized and raw one-bit covariance estimates, we find it to be critical for the Van Vleck and hybrid estimates.  We also find that performance can be quite sensitive to the level of loading.

This paper is organized as follows. In Section \ref{sec:population_model} we establish our array model. In Section \ref{sec:one_bit_quantization} we discuss the impact of one-bit-quantized data covariance estimates on the SINR. In Section \ref{sec:hybrid_cross_covariance}, we analyze a hybrid covariance estimation technique that uses both one-bit and full precision data. In Section \ref{sec:results} we discuss simulation results. Section \ref{sec:conclusion} is a brief summary of our work. Derivations are largely in the appendices.

\section{Population Model}
\label{sec:population_model}
We consider an array processing model with parameters $K$ (number of antenna elements), $L$ (number of time samples), and $M$ (number of interferers):
\begin{equation}
\label{array_model}
    Z = V\Sigma S^H + N,
\end{equation}
where $Z \in \mathbb{C}^{K \times L}$ is a complex Gaussian data matrix. $V \in \mathbb{C}^{K \times M}$ is the set of array response vectors to the $M$ interferers, with each entry assumed to be phase-only and therefore each column having norm $\sqrt{K}$. $\Sigma = \text{diag}\left(\sigma_1,...,\sigma_M\right) \in \mathbb{R}^{M \times M}$ is a diagonal matrix of root-mean square interference amplitudes, and $S$ is independent, complex Gaussian interference with rows $\mathbf{s}_l \sim \mathcal{CN}(\mathbf{0},I_M)$.  $N$ is complex additive white Gaussian noise (AWGN), with columns $\mathbf{n}_l \sim \mathcal{CN}(\mathbf{0},I_K)$. Note that the noise is normalized to unity.  The associated population covariance is
\begin{align}
\label{populationCovariance}
    R &= \mathbb{E}\left[\frac{1}{L}ZZ^H\right] \nonumber \\
    &= I + V \Sigma^2 V^H,
\end{align}
with sample covariance estimator
\begin{equation}
    \widehat{R} = \frac{1}{L} ZZ^H.
\end{equation}

\section{One-Bit Quantization}
\label{sec:one_bit_quantization}

Consider quantizing $Z$ to one real and one imaginary bit, i.e., 2 bits of complex data normalized to unity as per
\begin{align}
    \widetilde{Z} = [\widetilde{\mathbf{z}}_1,...,\widetilde{\mathbf{z}}_L] \equiv \frac{ \text{sgn}\left(\text{Re}(Z)\right) + j \cdot \text{sgn}\left(\text{Im}(Z)\right)}{\sqrt{2}},
\end{align}
where $\text{sgn}(\cdot)$ is applied separately to each entry of the matrix argument.  It is known \cite{VanVleckMiddleton} that if $\rho$ denotes the correlation between two real, jointly Gaussian random variables, then the correlation of the one-bit-quantized data is $\frac{2}{\pi}\arcsin(\rho)$. Applying this to the complex population covariance matrix $\widetilde{R}$ for the one-bit data yields
\begin{align}
    \widetilde{R} &= \mathbb{E}\left[\widetilde{\mathbf{z}}\,\widetilde{\mathbf{z}}^H\right] \nonumber \\
    &= \frac{2}{\pi} \left[\arcsin\left(\frac{\text{Re}(R)}{\beta}\right) + j \cdot \arcsin\left(\frac{\text{Im}(R)}{\beta}\right)\right],
    \label{Van_Vleck}
\end{align}
 where
 \begin{equation}
     \beta = 1 + \sum_{m=1}^{M} \sigma_m^2
 \end{equation} is the total average power seen by any element of the array due to the noise floor and the $M$ interferers. The arcsine function and expectation operator $\mathbb{E}[\cdot]$ are applied entrywise to the real and imaginary components of the matrix argument.  Thus, a correlation matrix built out of the one-bit samples is biased away from the true population correlation matrix, with higher correlation entries compressed more.  For small $x$, $\arcsin{x} \approx x$, so that smaller correlations are nearly preserved up to a multiplicative factor.  When most of the off-diagonal elements are small, the one-bit covariance keeps the spiked structure of $R$ but with weaker spikes (Section~\ref{sec:population_results}), so its nulls are shallower.

 We are interested in null depths and achievable SINRs under the various covariance estimates as a gauge of their practicality.  Here we consider the minimum variance distortionless response (MVDR) beamformer \cite[Sec.~6.2.1]{mvdr}, i.e.,
 \begin{equation}
 \label{mvdr}
     \widehat{\mathbf{w}} = \frac{\widehat{R}^{-1}\mathbf{v}_0}{\mathbf{v}_0^H \widehat{R}^{-1} \mathbf{v}_0},
 \end{equation}
where $\widehat{R}$ is the estimated covariance and $\mathbf{v}_0$ is the norm-$\sqrt{K}$ look vector to the signal of interest (SOI).  Note that by construction, $\widehat{\mathbf{w}}$ satisfies $\widehat{\mathbf{w}}^H \mathbf{v}_0 = 1$.  Thus, the output SINR is \cite[Ch.~11]{SINR}
 \begin{align}
 \label{sinr}
     \text{SINR} = \frac{\sigma_s^2}{\widehat{\mathbf{w}}^H R \widehat{\mathbf{w}}},
 \end{align}
 where $\sigma_s^2$ is the power of the SOI and $R$ is, strictly speaking, the covariance of the interference and noise only (i.e., excluding signal); the training data $Z$ in \eqref{array_model} are likewise taken to be signal-free.  The denominator is then the output power of the noise and interference after adaptive weighting, and is given by
 \begin{equation}
     \widehat{\mathbf{w}}^H R \widehat{\mathbf{w}} = \widehat{\mathbf{w}}^H \widehat{\mathbf{w}} + \widehat{\mathbf{w}}^H V \Sigma^2 V^H \widehat{\mathbf{w}},
 \end{equation}
 which consists of a residual white noise term and a residual interference term, respectively.  For the subsequent analysis, we make use of the following assumptions at various stages:

 \begin{enumerate}
     \item The array is large and the number of samples is much larger:
     \begin{equation}
         K \gtrsim 100, L \gg K. \label{large_array_assumption}
     \end{equation}
     In the subsequent SINR comparisons, we implicitly approximate the expected SINRs in terms of reciprocals of the expected residual interference and noise levels.  In particular, we have
     \begin{align}
         \mathbb{E}\left[\frac{1}{X}\right] = \frac{1}{\mathbb{E}[X]}\left(1 + O\left(\frac{\text{Var}(X)}{(\mathbb{E}[X])^2}\right)\right),
     \end{align}
     where $X = \widehat{\mathbf{w}}^H R \widehat{\mathbf{w}}$.
     As long as the inverse sample covariance estimates (including diagonal loading) remain well conditioned, the weights concentrate about their population values asymptotically in $L$. Thus, the output SINR also concentrates with a variance decreasing with increasing $L$, justifying the leading order approximation of the expectation of $X^{-1}$ above.

     \item  The interference array response vectors are mutually orthogonal to leading order:
      \begin{equation}
      \label{interference_orthogonality}
     \frac{1}{K} V^H V \approx I.
 \end{equation}
    For fixed, distinct angles of arrival, this becomes more reasonable as the number of antennas grows.

    \item \label{loud_interference_assumption} The interference covariance spikes are much louder than the noise, i.e.,
    \begin{equation}
    \label{strong_spikes}
        K \sigma_m^2 \gg 1.
    \end{equation}
    Note this is a statement about the interference after array gain, rather than the raw signal level at the array.

 \item Overlap between the SOI array response vector and the interference array responses is small. Defining
 \begin{equation}
    \boldsymbol{\eta} \equiv \frac{1}{K} V^H \mathbf{v}_0,
 \end{equation}
 this means
 \begin{equation}
     ||\boldsymbol{\eta}||^2 \ll 1.  \label{low_overlap_assumption}
 \end{equation}

 \item \label{small_correlation_assumption} The normalized off-diagonal correlations are small, i.e., $|R_{ij}|/\beta \ll 1$ for $i \neq j$, so that the arcsine law \eqref{Van_Vleck} can be linearized (Appendix \ref{sec:linearizing_one_bit_covariance}). This requires the interference power to be spread over many sources rather than dominated by a few, and underlies all of the one-bit and hybrid results.
 \end{enumerate}

\subsection{Population Covariance Results}
\label{sec:population_results}
We first consider the weights under the true population covariance $R$. Consider writing the SOI look vector in terms of the interference array response vectors $\mathbf{v}_m$ and a vector $\mathbf{v}_\perp$ with norm $\sqrt{K}$ that is orthogonal to the interference:
\begin{equation}
\label{v0_decomposition}
    \mathbf{v}_0 = \sum_{m=1}^{M} \eta_m \mathbf{v}_m + \sqrt{1 - ||\boldsymbol{\eta}||^2} \mathbf{v}_\perp.
\end{equation}
Note that by construction $\mathbf{v}_\perp$ is in the noise subspace of $R$, and therefore has eigenvalue $1$. The numerator of \eqref{mvdr} then becomes
\begin{equation}
    \inv{R} \mathbf{v}_0 = \sum_{m=1}^{M} \frac{\eta_m \mathbf{v}_m}{1 + K\sigma_m^2} + \sqrt{1 - ||\boldsymbol{\eta}||^2} \mathbf{v}_\perp,
\end{equation}
and the denominator is
\begin{align}
\label{v0_Rinv_v0}
    \mathbf{v}_0^H \inv{R} \mathbf{v}_0 = K - \sum_{m=1}^{M} \frac{|\eta_m|^2 K^2 \sigma_m^2}{1 + K\sigma_m^2}.
\end{align}
Thus, the residual noise satisfies
\begin{align}
\label{wng}
    \mathbf{w}^H \mathbf{w} &= \frac{1 - ||\boldsymbol{\eta}||^2 + \sum_{m=1}^{M} \frac{|\eta_m|^2}{\left(1 + K\sigma_m^2\right)^2}}{K \left(1 - \sum_{m=1}^{M} \frac{K \sigma_m^2 |\eta_m|^2}{1 + K\sigma_m^2 }\right)^2} \\
    &\approx \frac{1}{K(1 - ||\boldsymbol{\eta}||^2)} \label{wng_simplified}
\end{align}
for strong interference spikes. A similar calculation for the interference term yields
\begin{align}
    J &= \mathbf{w}^H V \Sigma^2 V^H \mathbf{w} = \frac{\sum_{m = 1}^{M} \frac{\sigma_m^2 |\eta_m|^2}{\left(1 + K\sigma_m^2\right)^2}}{\left(1 - \sum_{m=1}^{M} \frac{K \sigma_m^2 |\eta_m|^2}{1 + K\sigma_m^2}\right)^2} \label{unquantized_interference_residual}\\
    &\approx \frac{1}{K^2 (1 - ||\boldsymbol{\eta}||^2)^2} \sum_{m=1}^{M} \frac{|\eta_m|^2}{\sigma_m^2},
\end{align}
where again we assume strong spikes. Thus in principle, the less overlap there is between the array response vector and the interference array response vectors, the smaller the residual interference power can get after adaptation.

 In Appendix \ref{sec:linearizing_one_bit_covariance}, we show that linearizing the one-bit population covariance gives it a spiked covariance structure, making the first order effective one-bit covariance
 \begin{equation}
 \label{linearized_one_bit_covariance}
     \widetilde{R} \approx \alpha I + \frac{2}{\pi \beta} V \Sigma^2 V^H,
 \end{equation}
 where
 \begin{equation}
        \alpha \equiv 1-\frac{2}{\pi\beta}(\beta-1).
    \end{equation}
 Since the MVDR weights are invariant to an overall covariance scaling, we are free to work with $\frac{\widetilde{R}}{\alpha}$.  Define
 \begin{equation}
     c \equiv \frac{2}{\pi \alpha \beta}.
 \end{equation}
 When taking strong interference limits below, we also consider the one-bit spikes to remain strong, i.e., not just $K\sigma_m^2 \gg 1$ but also $c K \sigma_m^2 \gg 1$.  Then the one-bit noise term follows from \eqref{wng} after the replacement $K\sigma_m^2 \rightarrow cK\sigma_m^2$, so that its strong-spike limit is also given by \eqref{wng_simplified}. On the other hand, the interference residual becomes
  \begin{align}
     \widetilde{J} &= \widetilde{\mathbf{w}}^H V \Sigma^2 V^H \widetilde{\mathbf{w}} \\
     &= \frac{\sum_{m = 1}^{M} \frac{\sigma_m^2 |\eta_m|^2 }{\left(1 + c K\sigma_m^2\right)^2}}{\left(1 - ||\boldsymbol{\eta}||^2 + \sum_{m=1}^{M} \frac{|\eta_m|^2}{1 + cK\sigma_m^2}\right)^2}
     \label{one_bit_interference_residual}\\
     &\approx \frac{1}{c^2 K^2 \left(1 - ||\boldsymbol{\eta}||^2\right)^2} \sum_{m=1}^{M} \frac{|\eta_m|^2}{\sigma_m^2}
 \end{align}
for strong interference.

Substituting in the full, corresponding noise term \eqref{wng} and interference terms \eqref{unquantized_interference_residual} and \eqref{one_bit_interference_residual} into \eqref{sinr} and expanding to first order in $|\eta_m|^2$, we arrive at
\begin{align}
    \frac{\text{SINR}_\text{Unquantized}}{\text{SINR}_\text{1-Bit}} &\approx 1 + P, \label{one_bit_population_ratio} \\
    P &\equiv K \sum_{m=1}^{M} \frac{\sigma_m^2 |\eta_m|^2}{\left(1 + cK\sigma_m^2\right)^2}, \nonumber
\end{align}
where the unquantized residual and the difference between the noise terms are negligible. Here $P$ is the interference leaking through the shallower one-bit nulls, relative to the output noise, and is small only if each interferer's leakage is well below the noise. If all $cK\sigma_m^2 \gg 1$, then $P \approx (\pi/2-1)^2\beta^2 K^{-1}\sum_m |\eta_m|^2/\sigma_m^2$. For the trial of Section~\ref{sec:results}, $P \approx 0.64$, a 2.1 dB population loss (2.2 dB with the exact arcsine law).

\subsection{Sample Covariance Results}
 The story is different for the sample covariances. As shown in Appendix \ref{sec:finite_sample_wng_correction}, the expected white noise gain when using the sample covariance goes roughly as
 \begin{equation}
 \label{wng_finite_sample_correction}
     \mathbb{E}[\widehat{\mathbf{w}}^H\widehat{\mathbf{w}}] \approx \left(1+\frac{K-M-1}{L}\right) \frac{1}{K\left(1 - ||\boldsymbol{\eta}||^2\right)},
 \end{equation}
 which is not a big change for $\frac{L}{K} \gg 1$, justifying the sample-independent approximation (in this regime) of
 \begin{equation}
     \mathbb{E}[\widehat{\mathbf{w}}^H \widehat{\mathbf{w}}] \approx \frac{1}{K(1-||\boldsymbol{\eta}||^2)}.
 \end{equation}

 But the interference gain term \textit{does} limit performance when using the sample covariance. It is shown in Appendix \ref{sec:interference_gain_derivation} that under assumptions \eqref{large_array_assumption}, \eqref{interference_orthogonality}, and \eqref{low_overlap_assumption} the expected interference gain is
 \begin{align}
 \label{unquantized_sample_interference_gain}
     \mathbb{E}[\widehat{J}] &\approx \frac{1}{(1 - ||\boldsymbol{\eta}||^2)L} \sum_{m=1}^{M} \frac{\sigma_m^2}{1 + K \sigma_m^2},
 \end{align}
 where the population residual \eqref{unquantized_interference_residual}, which is small compared to this finite-sample term, is neglected.

If we again implicitly consider a Gaussian surrogate for the one-bit data with a population covariance given by \eqref{linearized_one_bit_covariance}, then a factor of $c$ is introduced that premultiplies $K \sigma_m^2$ in the denominator of the interference gain. With that in mind, the ratio of sample SINRs is then
\begin{align}
    & \frac{\mathbb{E}[\text{SINR}_\text{Unquantized}]}{\mathbb{E}[\text{SINR}_\text{1-Bit}]} \approx \frac{\frac{1}{K} + \frac{1}{L}\sum_{m=1}^{M} \frac{\sigma_m^2}{1+cK \sigma_m^2}}{\frac{1}{K} + \frac{1}{L}\sum_{m=1}^{M} \frac{\sigma_m^2}{1+K \sigma_m^2}} \nonumber \\
    &\approx 1 + \frac{K}{L}\sum_{m=1}^{M} \sigma_m^2 \left[\frac{1}{1 + c K\sigma_m^2} - \frac{1}{1 + K \sigma_m^2}\right],
\end{align}
where the common factor $\frac{1}{1 - ||\boldsymbol{\eta}||^2}$ of the noise and interference terms cancels, and where the second line follows from \eqref{large_array_assumption} (since $L \gg K > M$) and the associated result that the unquantized residual interference is thus small relative to the white-noise residual.  If all the effective one-bit spikes satisfy $c K\sigma_m^2 \gg 1$, this expression reduces to
\begin{equation}
\label{one_bit_sinr_ratio}
    \boxed{\frac{\mathbb{E}[\text{SINR}_\text{Unquantized}]}{\mathbb{E}[\text{SINR}_\text{1-Bit}]} \approx 1 + \left(\frac{\pi}{2} - 1\right) \beta\frac{M }{L}}.
\end{equation}

\subsection{Van Vleck Correction}

One can invert \eqref{Van_Vleck} entrywise through the so-called Van Vleck correction to arrive at an estimate of the unquantized covariance matrix. While purely one-bit samples cannot be used to obtain absolute channel power estimates, if channel powers are separately estimated, a diagonal matrix of \textit{estimated} powers
\begin{equation}
    \widehat{B} \equiv \text{diag}\left({\widehat{R}_{11}},...,\widehat{R}_{KK}\right)
\end{equation}
is available. Defining
\begin{equation}
    \widehat{\widetilde{R}} = \frac{1}{L} \widetilde{Z}\,\widetilde{Z}^H,
\end{equation}
this leads to the following correction to the estimator of the unquantized covariance matrix:
\begin{equation}
    \widehat{R}_\text{VV} = \widehat{B}^{\frac{1}{2}} \cdot \left[\sin\left(\frac{\pi}{2}\text{Re}(\widehat{\widetilde{R}})\right) + j \cdot \sin\left(\frac{\pi}{2}\text{Im}(\widehat{\widetilde{R}})\right)\right]  \cdot
    \widehat{B}^{\frac{1}{2}},
\end{equation}
where sine is applied entrywise to the real and imaginary components of the argument.

\begin{figure*}[!t]
\centering
\includegraphics[width=\textwidth]{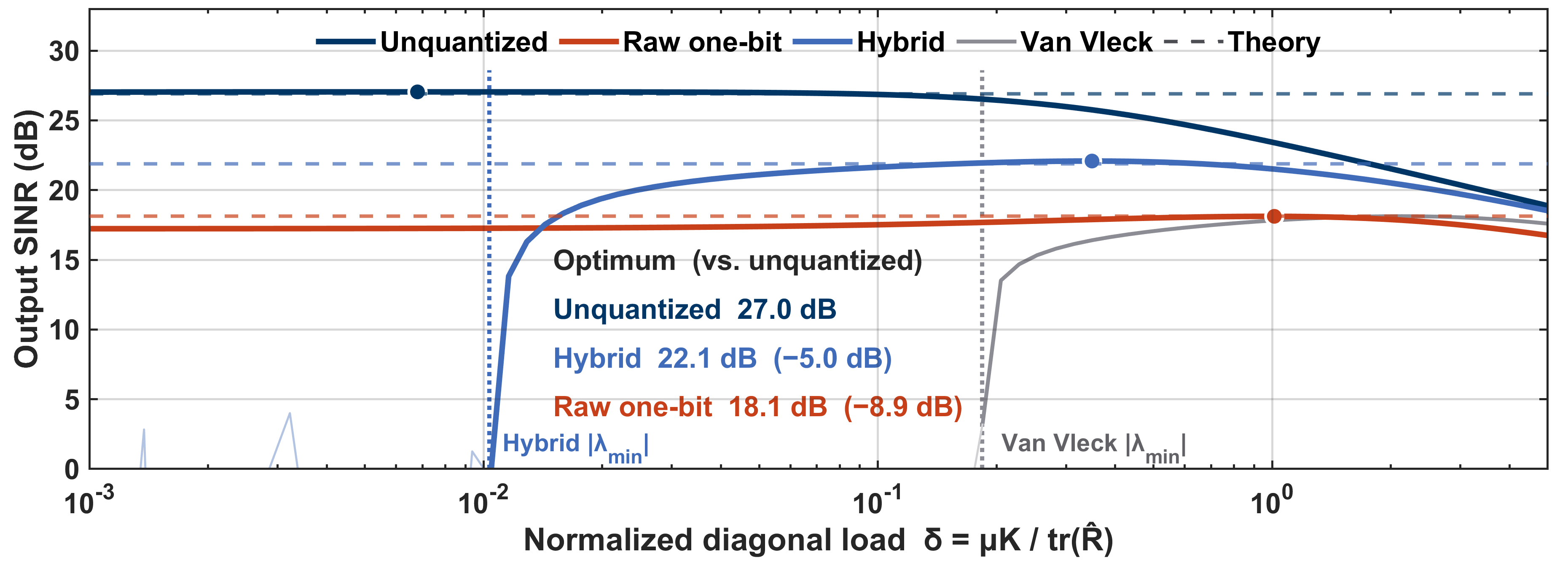}
\caption{Simulated SINR as a function of normalized diagonal load $\delta = \mu K / \text{tr}(\widehat{R})$ for the unquantized, one-bit (raw and Van Vleck-corrected), and hybrid covariance estimates, where $\mu$ is the diagonal load and $\widehat{R}$ is the corresponding estimate.  For this particular trial, the unquantized, raw one-bit, and hybrid schemes all approach their corresponding finite sample SINR approximations (dashed).  The minimum diagonal loads needed to achieve non-spurious SINR results for the Van Vleck and hybrid approaches are close in magnitude to the most negative eigenvalues of their respective covariance estimates (dotted vertical lines); below these loads, the curves are drawn faint. It is worth noting that small eigenvalues, whether positive or negative, can cause ill-conditioning of the covariance matrix estimates.  The difference is that for a PD matrix with small eigenvalues, any small diagonal load improves the conditioning of the matrix, while the condition number for a matrix with negative eigenvalues can degrade if an insufficient diagonal load is added, since large-magnitude negative eigenvalues can become very small upon loading.  That said, if a sufficiently large gap exists in the negative eigenvalues of the sample cross covariance, it is possible for optimal SINR performance to be achieved without a loading that forces PD, despite the non-physicality of those negative eigenvalues.  We have observed this behavior in some simulation trials for the hybrid covariance estimate, but not for the Van Vleck covariance.}
\label{fig:sinr_curves}
\end{figure*}

While it can be shown that $\widehat{R}_\text{VV}$ is a consistent estimator of $R$, the sine correction is not guaranteed to preserve positive definiteness.  Indeed, in simulation we often see small negative eigenvalues, so that the estimated covariance requires additional and more careful diagonal loading than its uncorrected one-bit counterpart.

We often see only marginal improvement in the corrected covariance SINR over the raw one-bit SINR after sweeping diagonal loads for both.  This is partly because the arcsine in the one-bit population covariance compresses the correlation magnitudes and effective spike sizes, mimicking the effects of regularization.
In practice, this means that if one is willing to use one-bit measurements to do adaptive beamforming, then forming the one-bit covariance may be sufficient.

\section{Hybrid Cross Covariance}
\label{sec:hybrid_cross_covariance}
 Rather than a covariance built fully out of one-bit samples, one can construct a symmetrized \textit{cross}-covariance between the one-bit data and the unquantized data.  This hybrid approach sacrifices some of the computational savings afforded by pure sign quantization, but also achieves higher SINRs. We define the cross covariance as
\begin{equation}
\label{sample_cross_covariance}
    \widehat{R}_h = \frac{1}{2L} \left(Z \widetilde{Z}^H + \widetilde{Z} Z^H\right),
\end{equation}
with mean
\begin{equation}
    \mathbb{E}\left[\widehat{R}_h\right] = \sqrt{\frac{2}{\pi \beta}}R.
    \label{hybrid_mean}
\end{equation}
This follows exactly from the Bussgang theorem \cite{BussgangTheorem} for Gaussian data with equal element powers.

Although a typical cross-covariance does not have the same properties as a true covariance, here the expected cross-covariance is a rescaled true population covariance, so the hybrid estimator achieves asymptotically optimal SINRs. However, with finite sampling, its performance tends to sit in between that of the unquantized and the one-bit sample covariances.  Additionally, though manifestly Hermitian, the hybrid estimator is typically not positive definite in practice and requires careful diagonal loading.

In Appendix \ref{sec:hybrid_interference_gain_derivation} we show that under assumptions \eqref{strong_spikes} and \eqref{low_overlap_assumption}, as well as additional restrictions on the magnitude of the diagonal load relative to the spike eigenvalues and the finite sample error, the expected interference residual for the hybrid sample covariance is approximately
\begin{equation}
    \mathbb{E}[\widehat{J}_h] \approx \frac{1 + \frac{\beta(\pi -2)}{8}}{(1 - ||\boldsymbol{\eta}||^2)L} \sum_{m=1}^{M} \frac{\sigma_m^2}{1+K\sigma_m^2},
\end{equation}
an inflation factor of roughly $1 + 0.14\beta$ over the residual interference in the unquantized scheme.  The noise residual follows from \eqref{hybrid_noise_residual}, which reduces to \eqref{wng_simplified} in the typical large-sample regime (provided the diagonal load scales in such a way as to prevent the interference magnitude from inflating the term on the RHS of \eqref{hybrid_noise_residual}).
For $M$ interferers satisfying $K \sigma_m^2 \gg 1$, the analogous result to \eqref{one_bit_sinr_ratio} is then
\begin{equation}
\label{hybrid_sinr_ratio}
    \boxed{\frac{\mathbb{E}[\text{SINR}_\text{Unquantized}]}{\mathbb{E}[\text{SINR}_h]} \approx 1 + \left(\frac{\pi - 2}{8}\right)\beta\frac{M }{L}}.
\end{equation}
Thus, the excess interference coefficients for the one-bit and hybrid cases differ by a factor of 4 in the limit of strong interference.
To leading order, the finite-sample noise and interference terms share a factor of $\frac{1}{1 - ||\boldsymbol{\eta}||^2}$, which cancels in the SINR ratios. Geometric dependence remains through finite-spike and higher-order corrections, and through the population residuals neglected here: \eqref{unquantized_interference_residual} is negligible, \eqref{one_bit_interference_residual} adds $P$ of \eqref{one_bit_population_ratio} to the one-bit excess, and the hybrid has none, since \eqref{hybrid_mean} is proportional to $R$.
As with the Van Vleck correction, we may choose to apply a power correction to the hybrid estimator. At the population level, this amounts to multiplication by a scalar and therefore leaves SINRs unchanged, while at the finite but sufficient sample level, this is just a perturbation and does not seem to produce meaningfully different SINRs. The hybrid results in Section \ref{sec:results} use $\widehat{R}_h$ without this correction.

\section{Results}
\label{sec:results}
For the simulation trial shown, we select a uniform linear half-wavelength antenna array of size $K = 512$ with $L = 25\cdot K = 12800$ samples.  $M=100$ interferers with powers selected uniformly at random from 0 to 20 dB are generated with azimuth angles $\theta \sim \mathcal{U}(0^\circ,360^\circ)$ and elevation angles $\phi \sim \mathcal{U}(30^\circ,90^\circ)$; any interferer within $20/K$ in direction cosine (about $2^\circ$) of the broadside look direction or within $4/K$ of another interferer is discarded and replaced by a new random draw. The SOI power is $\sigma_s^2 = 1$, and $\beta \approx 2034$.  Simulated SINR results are shown for one random trial in Fig.~\ref{fig:sinr_curves} for the unquantized, one-bit (raw and Van Vleck-corrected), and hybrid covariance estimates.  The given trial satisfies $||\boldsymbol{\eta}||^2 = 0.0033$, and $\frac{1}{K}V^H V$ is close to identity, with off-diagonal entries no larger than $0.13$ in magnitude, so that the geometry satisfies assumptions \eqref{interference_orthogonality} and \eqref{low_overlap_assumption}.  Although the weakest (near 0 dB) interferers have $cK\sigma_m^2 \approx 0.5$ rather than $cK\sigma_m^2 \gg 1$, the unquantized, hybrid, and raw one-bit SINRs all peak within 0.2 dB of their predictions. For the raw one-bit estimate, the peak occurs at a load comparable to its weakest spikes, beyond the small-load approximation; the resulting gain roughly offsets the neglected population term $P$.

For each estimator, the diagonal load level is varied from $10^{-3}$ to 5 times the trace of the corresponding covariance estimate divided by $K$. The finite-sample SINR approximations derived above, evaluated before taking the strong-spike limits, are also shown for comparison.  Vertical lines are included to denote the normalized magnitude of the most negative eigenvalue of both the Van Vleck one-bit covariance and the hybrid covariance.  Note that the SINR behavior of these two estimators tends to be spurious below those values, as a diagonal load with magnitude less than that of the most negative eigenvalue is insufficient for alleviating poor matrix conditioning.  In that region, the matrix is not interpretable as a positive-definite covariance estimate.

To test the boxed approximations within their stated regime, we also ran a Monte Carlo study with $K = 512$, $M = 32$ equal-power interferers, $L/K \in \{20, 40, 80\}$, and $\beta$ from 100 to 3000 (60 trials per point, with each estimator at its best-on-average load). The measured excess penalties $\mathbb{E}[\text{SINR}_\text{Unquantized}]/\mathbb{E}[\text{SINR}] - 1$ of the raw one-bit and hybrid estimators agree with \eqref{one_bit_sinr_ratio} and \eqref{hybrid_sinr_ratio} to within about 10\% at all but the smallest penalties, and their ratio is 3.9 (range 3.7--4.5), consistent with the predicted factor of 4.

All of the curves exhibit sensitivity to the swept diagonal load.  For a real system implementing any of these covariance estimation schemes to work optimally, the load parameter must be chosen carefully---too low and some of the schemes fail completely, too high and the weights collapse to a passive beamformer.  While we have come across existing methods for setting diagonal loading in the unquantized case, there is no guarantee that such approaches work for the one-bit and hybrid estimators.  Further algorithmic investigation into load selection is required.

\section{Conclusion}
\label{sec:conclusion}
In this work, we considered spatial covariance estimation using complex data that has been quantized to one complex bit.  We presented an approach to build a covariance matrix fully out of the one-bit data, as well as a technique used to correct it conditioned on jointly Gaussian data.  The second approach uses one copy of the one-bit data and one copy of the full precision data to achieve greater SINRs than the fully one-bit case, but requires more computation.  We noted that careful diagonal loading is required for these approaches to achieve meaningful SINRs.
\appendices
\section{Linearizing the One-Bit Covariance}
\label{sec:linearizing_one_bit_covariance}
We would like to approximate the one-bit population covariance in \eqref{Van_Vleck} as a spiked covariance. To that end, we split up the correlation matrix $\frac{R}{\beta}$ into the sum of an identity matrix and an off-diagonal matrix, which enables us to ``distribute'' the entrywise arcsin over the argument:
 \begin{align}
     \widetilde{R}_{ij} &= \frac{2}{\pi} \arcsin{\left(\delta_{ij} + \frac{R_{ij}}{\beta}-\delta_{ij}\right)} \nonumber \\
     &= \delta_{ij} + \frac{2}{\pi} \arcsin{\left(\frac{\delta_{ij} + \sum_{m=1}^{M}V_{im} \sigma_m^2 V_{jm}^*}{\beta}-\delta_{ij}\right)} \nonumber \\
     &\approx \delta_{ij} + \frac{2}{\pi}\left(\frac{\delta_{ij} + \sum_{m=1}^{M}V_{im} \sigma_m^2 V_{jm}^*}{\beta} - \delta_{ij}\right),
 \end{align}
 where $\delta_{ij}$ is the Kronecker delta, and in going to the third line we use the Taylor series $\arcsin(x) = x + \frac{x^3}{6} + O(x^5)$ about $x = 0$.  To first order in the off-diagonal entries of the population correlation matrix, we have
 \begin{align}
     \widetilde{R} &\approx \left(1 + \frac{2}{\pi \beta}\left(1 - \beta\right)\right)I + \frac{2}{\pi \beta} V \Sigma^2 V^H \nonumber \\
     &= \alpha I + \frac{2}{\pi \beta}V \Sigma^2 V^H,
 \end{align}
 with $\alpha \equiv 1 + \frac{2}{\pi \beta}(1-\beta) \approx 0.36$ for large $\beta$.  Note that this is a spiked covariance matrix with spike eigenvalues $\lambda_m = \alpha(1 + cK\sigma_m^2), m \in \{1,...,M\}$ and bulk eigenvalues $\lambda_j = \alpha, j \in \{M+1,...,K\}$.

 Since the MVDR is invariant to a global scaling on the covariance, we are free to consider instead the matrix
 \begin{equation}
     \widetilde{R} \approx I + c V \Sigma^2 V^H,
 \end{equation}
 which clarifies the comparison to the unquantized population covariance: to first order, quantizing the data shrinks spike levels by the factor $c \approx \frac{7}{4\beta}$ for large $\beta$.

\section{Finite Sample Correction to White Noise Gain}
\label{sec:finite_sample_wng_correction}
From \eqref{mvdr}, we have
\begin{equation}
    \widehat{\mathbf{w}}^H \widehat{\mathbf{w}} = \frac{\mathbf{v}_0^H \widehat{R}^{-2}\mathbf{v}_0}{|\mathbf{v}_0^H \widehat{R}^{-1} \mathbf{v}_0|^2}.
\end{equation}

When $R = I$ (the noise-only case), Reed, Mallett, and Brennan \cite{rmb} show that
\begin{equation}
    \frac{1}{K\widehat{\mathbf{w}}^H \widehat{\mathbf{w}}} \sim \text{Beta}(L-K+2,K-1).
\end{equation}

 Thus,
 \begin{align}
 \label{finite_sample_wng_isotropic}
     \mathbb{E}[\widehat{\mathbf{w}}^H \widehat{\mathbf{w}}] &= \frac{L}{K(L-K+1)} \\
     &\approx \frac{1}{K} \left(1 + \frac{K-1}{L}\right),
 \end{align}
 i.e., the finite sample correction for the noise is roughly $1 + \frac{K-1}{L}$.

 Given the decomposition \eqref{v0_decomposition}, in the limit of strong interference, the MVDR weights are mostly spanned by $\mathbf{v}_\perp$, so that to satisfy the distortionless constraint we must have
 \begin{equation}
     \widehat{\mathbf{w}}^H \widehat{\mathbf{w}} \approx \frac{1}{K(1-||\boldsymbol{\eta}||^2)}.
 \end{equation}
 Now, adding in the assumption of low overlap between SOI and interference, we expect that the corrected, approximate white-noise gain term for the spiked covariance matrix goes as
 \begin{equation}
     \mathbb{E}[\widehat{\mathbf{w}}^H \widehat{\mathbf{w}}] \approx \frac{1}{K(1-||\boldsymbol{\eta}||^2)} \left(1 + \frac{K - M - 1}{L}\right),
 \end{equation}
 where we heuristically subtract $M$ from $K - 1$ due to the $M$ strong non-noise eigenvalues that are approximately removed from the degrees of freedom when inverting the sample covariance.
 This count is consistent with \cite{rmb}, whose result holds for any $R$:
 \begin{align*}
     \mathbb{E}[\widehat{\mathbf{w}}^H R \widehat{\mathbf{w}}] &= \frac{L}{(L-K+1)\mathbf{v}_0^H R^{-1} \mathbf{v}_0} \\
     &\approx \frac{1}{K(1 - ||\boldsymbol{\eta}||^2)}\left(1 + \frac{K-1}{L}\right)
 \end{align*}
 by \eqref{large_array_assumption} and \eqref{v0_Rinv_v0} with strong spikes, and subtracting the interference residual \eqref{unquantized_sample_interference_gain}, approximately $\frac{M}{(1 - ||\boldsymbol{\eta}||^2)KL}$ under \eqref{strong_spikes}, recovers the result above.  This correction for the spiked covariance matrix stems from the assumption---in the strong interference and low overlap regime---that the look vector lies predominantly in the noise subspace.
 Practically speaking, the more interference sources in the environment, the less this result applies, as the expected overlap between $\mathbf{v}_0$ and $V$ begins to increase.

\section{Interference Gain Derivations}
\label{sec:interference_gain_derivation}
We would like to compute the unquantized sample covariance interference gain
\begin{equation}
\label{sample_interference_gain_expression}
    \widehat{J} = \frac{{\mathbf{v}_0}^H \widehat{R}^{-1} V \Sigma^2 V^H \widehat{R}^{-1} \mathbf{v}_0}{|\mathbf{v}_0^H \widehat{R}^{-1} \mathbf{v}_0|^2}.
\end{equation}
Define $\mathbf{u}_m = \frac{\mathbf{v}_m}{\sqrt{K}}, m \in \{1,...,M\}$, and $\mathbf{u}_0 = \frac{\mathbf{v}_\perp}{\sqrt{K}}$, with $U_J = [\mathbf{u}_1,...,\mathbf{u}_M] $ and $U = [\mathbf{u}_0, U_J]$. Additionally, let $\lambda_m = 1+ K\sigma_m^2$ and $\Lambda \equiv \text{diag}(\lambda_1,...,\lambda_M)$. By \eqref{v0_decomposition}, $\mathbf{v}_0 = \sqrt{K}\left(\sqrt{1 - ||\boldsymbol{\eta}||^2}\,\mathbf{u}_0 + U_J \boldsymbol{\eta}\right)$. Under \eqref{interference_orthogonality} the columns of $U$ are orthonormal, and since $\mathbf{u}_0$ is orthogonal to the interference, $R\mathbf{u}_0 = \mathbf{u}_0$ for any $\boldsymbol{\eta}$.  Assume $\widehat{R}$ has diagonal load level $\mu$, so that $\widehat{R}(\mu) = \widehat{R} + \mu I$.

Consider now the complete unitary basis $\mathcal{U} = [U, U_\perp]$. We then partition $\widehat{R}(\mu)$ as
\begin{equation}
    \mathcal{U}^H \widehat{R}(\mu)\mathcal{U} = \begin{pmatrix}
        A & B \\ B^H & D
    \end{pmatrix},
\end{equation}
where
\begin{equation}
    A \equiv \begin{pmatrix}
        r_{00} + \mu & \mathbf{g}^H \\ \mathbf{g} & \widehat{\Lambda} + \mu I
    \end{pmatrix},
\end{equation}
with $r_{00} = \mathbf{u}_0^H \widehat{R} \mathbf{u}_0$, $\mathbf{g} = U_J^H \widehat{R} \mathbf{u}_0$, and $\widehat{\Lambda} = U_J^H \widehat{R} U_J$. The block $A$ represents SOI-interference coupling and $B$ represents finite-sample coupling between directional sources (interference and SOI) and noise; since $U^H R U_\perp = 0$, $B$ is purely sampling error. The $U$-space component of the inverse covariance then corresponds to the top left block of $\inv{\left(\mathcal{U}^H \widehat{R}(\mu)\mathcal{U}\right)}$, which is given by the Schur complement $\inv{\left(A - B \inv{D} B^H\right)}$ of $A$.  Now, since $B \inv{D} B^H$ is second order in the finite sample error that couples $U$ and $U_\perp$, we neglect that term under the assumption $L \gg K$.  Thus, in the $U$-subspace, the MVDR weights go as
\begin{equation}
    U^H\mathbf{\widehat{w}} \approx \frac{1}{\sqrt{K(1 - ||\boldsymbol{\eta}||^2)}} \begin{pmatrix} 1 \\ -(\widehat{\Lambda} + \mu I)^{-1} \mathbf{g}\end{pmatrix},
\end{equation}
to first order in $||\boldsymbol{\eta}||^2$, plus a term $\frac{1 + \mu}{\sqrt{K}(1 - ||\boldsymbol{\eta}||^2)}(\widehat{\Lambda} + \mu I)^{-1}\boldsymbol{\eta}$ in the $U_J$ components due to $U_J\boldsymbol{\eta}$ in $\mathbf{v}_0$, so that the interference residual is
\begin{equation}
    \widehat{J} \approx \frac{1}{1 - ||\boldsymbol{\eta}||^2} \mathbf{g}^H (\widehat{\Lambda} + \mu I)^{-1} \Sigma^2 (\widehat{\Lambda} + \mu I)^{-1} \mathbf{g}
\end{equation}
plus a cross term linear in $\mathbf{g}$ and, with $\widehat{\Lambda} \rightarrow \Lambda$, the loaded population residual
\begin{equation*}
    J(\mu) \approx \frac{1}{(1 - ||\boldsymbol{\eta}||^2)^2} \sum_{m=1}^{M} \frac{\sigma_m^2 |\eta_m|^2 (1 + \mu)^2}{(\lambda_m + \mu)^2},
\end{equation*}
which reduces to \eqref{unquantized_interference_residual} at $\mu = 0$ for strong spikes. Since $\mathbb{E}[\mathbf{g}] = U_J^H R \mathbf{u}_0 = \mathbf{0}$, the cross term vanishes in expectation, and $J(\mu)$ is negligible relative to the finite-sample term below when $L(1 + \mu)^2 |\eta_m|^2 \ll \lambda_m$, so we drop both.

Now, with $\widehat{R} = R + E$ for some generic error matrix $E$, we have
\begin{align}
    \widehat{\Lambda} &= U_J^H \widehat{R} U_J = \Lambda + U_J^H E U_J.
\end{align}
Writing $F = U_J^H E U_J$, we have
\begin{equation}
    \inv{(\Lambda + \mu I + F)} = \inv{(I + \inv{(\Lambda + \mu I)} F)} \inv{(\Lambda + \mu I)}.
\end{equation}
Whenever $||\left(\Lambda + \mu I\right)^{-1} F|| \ll 1$---that is, when $||F|| \ll 1 + K \text{min}_m(\sigma_m^2) + \mu$---a Neumann expansion yields
\begin{equation}
    \inv{(\Lambda + \mu I + F)} \approx \inv{(\Lambda + \mu I)}.
\end{equation}
This condition holds as long as the diagonal load is set so that the finite sample error is small relative to the minimal loaded interference subspace eigenvalue. We thus have
\begin{equation}
\label{j_of_g2}
    \widehat{J} \approx \frac{1}{1 - ||\boldsymbol{\eta}||^2} \sum_{m=1}^{M} \frac{\sigma_m^2 |g_m|^2}{\left(\lambda_m + \mu\right)^2}.
\end{equation}
Therefore, to get at $\mathbb{E}[\widehat{J}]$, we need to look at the expected couplings $\mathbb{E}|g_m|^2$.

The explicit expression for $g_m$ is
\begin{equation}
    g_m = \frac{1}{L} \sum_{l=1}^{L} \mathbf{u}_m^H \mathbf{z}_l (\mathbf{u}_0^H \mathbf{z}_l)^*.
\end{equation}
Since $\mathbf{u}_m^H R \mathbf{u}_0 = 0$ exactly, the cross-sample terms in $\mathbb{E}|g_m|^2$ vanish, leading to
\begin{align}
\label{E_of_g2}
    \mathbb{E} |g_m|^2 &= \frac{1}{L} \mathbb{E}\left[|\mathbf{u}_m^H \mathbf{z}_l|^2|(\mathbf{u}_0^H \mathbf{z}_l)^*|^2\right] \nonumber \\
    &= \frac{1}{L} (\mathbf{u}_m^H R \mathbf{u}_m) (\mathbf{u}_0^H R \mathbf{u}_0)\\
    &\approx \frac{\lambda_m}{L},
\end{align}
using Isserlis' theorem \cite{isserlis} and $\mathbf{u}_0^H R \mathbf{u}_0 = 1$. In expectation, we arrive at
\begin{align}
\label{unquantized_interference_gain_derived}
    \mathbb{E}[\widehat{J}] &\approx \frac{1}{(1 - ||\boldsymbol{\eta}||^2)L} \sum_{m=1}^{M} \frac{\sigma_m^2 \lambda_m}{\left(\lambda_m + \mu\right)^2} \\
    &\approx \frac{1}{(1 - ||\boldsymbol{\eta}||^2)L} \sum_{m=1}^{M} \frac{\sigma_m^2}{\lambda_m + \mu}
\end{align}
for small diagonal load, i.e., $\mu \ll 1 + K\sigma_m^2$.  The more pressing reason for the small loading condition is that a large diagonal load leads to weights that reproduce a passive beamformer. In this regime, \eqref{unquantized_interference_gain_derived} incorrectly predicts zero residual interference power, because it omits $J(\mu)$, which grows with the load and tends to the passive-beamformer residual $\sum_{m} \sigma_m^2 |\eta_m|^2$ (to leading order in $||\boldsymbol{\eta}||^2$) as $\mu \rightarrow \infty$, so that the condition for dropping it fails.
\subsection{One-Bit}
While the one-bit samples are not Gaussian, we consider a Gaussian surrogate for the quantized data with a population covariance matching the linearized arcsin covariance. This allows us to make use of the result above to approximate the interference gain.  In particular, using the linearized arcsin model with diagonal load $\widetilde{\mu}$, we scale the $K \sigma_m^2$ term in \eqref{unquantized_interference_gain_derived} by $c$, which yields the one-bit sample covariance interference gain approximation
\begin{align}
    \mathbb{E}[\widehat{\widetilde{J}}] &\approx \frac{1}{(1 - ||\boldsymbol{\eta}||^2)L} \sum_{m=1}^{M} \frac{\sigma_m^2 \left(1 + c K \sigma_m^2\right)}{\left(1+c K\sigma_m^2 + \widetilde{\mu}\right)^2} \\
    &\approx \frac{1}{(1 - ||\boldsymbol{\eta}||^2)L} \sum_{m=1}^{M} \frac{\sigma_m^2}{1+cK\sigma_m^2 + \widetilde{\mu}}.
\end{align}
Note that this assumes that the one-bit data follows an effective spiked covariance model in order to capitalize on the form of \eqref{unquantized_interference_gain_derived}. Since $\mathbb{E}|g_m|^2$ involves fourth-order moments, which the surrogate does not reproduce, this is a modeling approximation beyond the assumptions of Section \ref{sec:one_bit_quantization}.

\subsection{Hybrid}
\label{sec:hybrid_interference_gain_derivation}
This derivation follows similar steps to the unquantized case, but the hybrid sample covariance now has a different structure. In particular, defining $c_h \equiv \sqrt{\frac{2}{\pi \beta}}$ and using the Bussgang decomposition \cite{BussgangTheorem} $\widetilde{\mathbf{z}}_l = c_h \mathbf{z}_l + \mathbf{q}_l$ for the one-bit data, where the quantization noise $\mathbf{q}_l$ is uncorrelated with $\mathbf{z}_l$, we arrive at
\begin{equation}
    \widehat{R}_h(\mu_h) = c_h\widehat{R} + \Delta R_h + \mu_h I,
\end{equation}
where
\begin{equation}
    \Delta R_h \equiv \frac{1}{2L} \sum_{l=1}^{L} \left(\mathbf{z}_l \mathbf{q}_l^H + \mathbf{q}_l \mathbf{z}_l^H\right),
\end{equation}
and $\mu_h$ is the diagonal load chosen for the hybrid sample cross-covariance and is in general different from $\mu$.
We will work with the rescaled cross covariance
\begin{align}
    \bar{R}_h \equiv c_h^{-1} \widehat{R}_h = \widehat{R} + \overline{\Delta R}_h + \bar{\mu}_h I,
\end{align}
where each overbar denotes scaling of the barred quantity by $\frac{1}{c_h}$. If we apply the same perturbative approach used in the unquantized case above, then the effective couplings now include both the error from the unquantized covariance \textit{and} the error matrix resulting from the Bussgang decomposition. They take the form
\begin{align}
\label{hybrid_couplings}
    g_m^{(h)} = g_m + \mathbf{u}_m^H \overline{\Delta R}_h \mathbf{u}_0.
\end{align}
In order to obtain a tractable model of the residual interference, we close the necessary fourth-order moments by factoring the mixed moments of $\mathbf{q}_l$ and $\mathbf{z}_l$, effectively treating the two as independent; note that this is a stronger assumption than the uncorrelatedness that the Bussgang theorem provides.  This assumption implies that
\begin{equation}
    \mathbb{E}[g_m^* \mathbf{u}_m^H \overline{\Delta R}_h \mathbf{u}_0] \approx 0.
\end{equation}
Using this approximate independence, the interference gain term
becomes
\begin{align}
    \mathbb{E}[\widehat{J}_h] \approx \frac{1}{1 - ||\boldsymbol{\eta}||^2} \sum_{m=1}^{M} \sigma_m^2 \frac{\mathbb{E}|g_m|^2 + \mathbb{E}|\mathbf{u}_m^H \overline{\Delta R}_h \mathbf{u}_0|^2}{\left(\lambda_m + \bar{\mu}_h\right)^2}.
\end{align}
We already know $\mathbb{E}|g_m|^2$ from \eqref{E_of_g2}, so only $\mathbb{E}|\mathbf{u}_m^H \overline{\Delta R}_h \mathbf{u}_0|^2$ remains to be calculated. To that end, we know
\begin{align}
    \mathbf{u}_m^H \overline{\Delta R}_h \mathbf{u}_0 &= \frac{1}{2L c_h} \sum_{l=1}^{L} \left(\mathbf{u}_m^H \mathbf{z}_l \mathbf{q}_l^H \mathbf{u}_0 + \mathbf{u}_m^H \mathbf{q}_l \mathbf{z}_l^H \mathbf{u}_0\right).
\end{align}
Again invoking the approximated independence between $\mathbf{q}_l$ and $\mathbf{z}_l$, as well as independence across time samples, we get
\begin{align}
    \mathbb{E} |\mathbf{u}_m^H \overline{\Delta R}_h \mathbf{u}_0|^2 &\approx \frac{1}{4 L c_h^2} [(\mathbf{u}_m^H R \mathbf{u}_m)(\mathbf{u}_0^H R_q \mathbf{u}_0) \nonumber \\
    &+ (\mathbf{u}_m^H R_q \mathbf{u}_m)(\mathbf{u}_0^H R \mathbf{u}_0)] \nonumber \\
    &\approx \frac{\beta (\pi - 2)(1 + \lambda_m)}{8L},
\label{expected_coupling_perturbation}
\end{align}
where we use $\mathbf{u}_m^H R \mathbf{u}_m = \lambda_m = 1 + K\sigma_m^2$ and $\mathbf{u}_0^H R \mathbf{u}_0 = 1$. Here $R_q \equiv \widetilde{R} - c_h^2 R \approx \left(1-\frac{2}{\pi}\right)I$ under the first-order arcsine approximation from before.
Thus we arrive at
\begin{align}
    \mathbb{E}[\widehat{J}_h] &\approx \frac{1}{(1 - ||\boldsymbol{\eta}||^2)L}\sum_{m=1}^{M} \frac{\sigma_m^2 \left(\lambda_m + \frac{\beta (\pi - 2)(1 + \lambda_m)}{8}\right)}{(\lambda_m + \bar{\mu}_h)^2}  \nonumber \\
    &\approx \frac{1}{(1 - ||\boldsymbol{\eta}||^2)L} \sum_{m=1}^{M}  \frac{\sigma_m^2}{\lambda_m + \bar{\mu}_h}\left(1 + \frac{\pi-2}{8}\beta\right),
\end{align}
where we again assume $K\sigma_m^2 \gg 1$ and $\bar{\mu}_h \ll 1 + K\sigma_m^2$.

The same coupling model can be used to derive the white-noise residual term under the hybrid scheme.  The couplings still take the form of \eqref{hybrid_couplings}, but now $m$ is allowed to range from $1$ to $K-1$, since all of the modes other than $\mathbf{u}_0$ are coupled when computing $\widehat{\mathbf{w}}^H \widehat{\mathbf{w}}$. Equation \eqref{expected_coupling_perturbation} holds as well if we allow $\lambda_m$ to label the population eigenvalues associated to the interference \textit{and} to the noise.  To leading order, we have
\begin{equation}
    \widehat{\mathbf{w}}_h^H \widehat{\mathbf{w}}_h \approx \frac{1}{K(1 - ||\boldsymbol{\eta}||^2)} \left(1 + \sum_{m=1}^{K-1} \frac{|g_m^{(h)}|^2}{(\lambda_m + \bar{\mu}_h)^2}\right),
\end{equation}
which, writing $\nu_m \equiv L\,\mathbb{E}|g_m^{(h)}|^2 \approx \lambda_m + \frac{\beta(\pi-2)}{8}(1+\lambda_m)$ from \eqref{E_of_g2} and \eqref{expected_coupling_perturbation}, leads to
\begin{align}
\mathbb{E}[\widehat{\mathbf{w}}_h^H \widehat{\mathbf{w}}_h]
&\approx \frac{1}{K(1 - ||\boldsymbol{\eta}||^2)} \biggl(1 + \sum_{m=1}^{M} \frac{\nu_m}{L(\lambda_m + \bar{\mu}_h)^2} \nonumber \\
&\quad {}+ \frac{(K-M-1)\gamma_h}{L(1+\bar{\mu}_h)^2}\biggr),
\end{align}
where $\gamma_h \equiv 1 + \frac{\beta(\pi-2)}{4}$ is the value of $\nu_m$ for the noise modes ($\lambda_m = 1$). In the strong interference spike limit, the sum over interferers vanishes, leading to
\begin{equation}
\label{hybrid_noise_residual}
    \mathbb{E}[\widehat{\mathbf{w}}_h^H \widehat{\mathbf{w}}_h] \approx \frac{1}{K(1-||\boldsymbol{\eta}||^2)} \left(1 + \frac{(K - M - 1)\gamma_h}{L(1 + \bar{\mu}_h)^2}\right).
\end{equation}
The quantization contribution to this residual noise term is negligible provided $L(1 + \bar{\mu}_h)^2 \gg (K - M - 1)\gamma_h$, which is also roughly the condition for the first-order expansion behind the preceding expressions to hold over the noise modes. For the trial of Section \ref{sec:results}, $\gamma_h \approx 582$, so this requires $1 + \bar{\mu}_h \gg 4.3$, which is compatible with $\bar{\mu}_h \ll 1 + K\sigma_m^2$ since $1 + K\sigma_m^2 \geq 513$. Setting the quantization terms to zero ($\gamma_h \rightarrow 1$) without loading ($\bar{\mu}_h = 0$) recovers \eqref{wng_finite_sample_correction}, independently supporting the $K - M - 1$ count of Appendix \ref{sec:finite_sample_wng_correction}.

\section*{Acknowledgment}
We would like to thank Helmuth Naumer for his feedback on the manuscript, as well as helpful discussions on the mathematics contained within.  We are also grateful to David Browne for his constant suggestions and guidance on the project.

\end{document}